\documentclass{svproc}
\usepackage{graphicx}
\usepackage{mathtools}
\usepackage{amsmath,amsfonts,amssymb}
\usepackage{tikz}
\usetikzlibrary {arrows.meta,bending,positioning}
\usepackage{tikz-dimline}
\usepackage{float}
\usepackage{subcaption}
\usepackage{bm}
\usepackage{multirow}
\usepackage{siunitx}
\usepackage{comment}
\usepackage{listings}
\usepackage{hyperref}
\hypersetup{pdftex=true, colorlinks=true, breaklinks=true, linkcolor=blue, menucolor=blue, citecolor=blue, urlcolor=blue}
\usepackage{xcolor}
\definecolor{beaublue}{rgb}{0.94, 0.97, 1.0}
\usepackage{url}

\begin{document}
	\mainmatter              
	\title{Topology optimization of contact-aided compliant mechanisms for tracing multi-kink paths}
	
	\titlerunning{Topology optimization of CCMs for tracing multi-kink paths}
	
	\author{Prabhat Kumar\inst{1,}\inst{2,}\inst{3}, Roger A Sauer\inst{4,}\inst{5,}\inst{6} \and
		Anupam Saxena\inst{7}}
	
	\authorrunning{Kumar et. al.}
	\institute{Department of Mechanical and Aerospace Engineering,\\
		Indian Institute of Technology Hyderabad,
		Telangana 502285, India  \\ \and
		Department of Computational Engineering,\\
		Indian Institute of Technology Hyderabad,
		Telangana 502285, India  \\ \and
		Department of Engineering Science,\\
		Indian Institute of Technology Hyderabad,
		Telangana 502285, India  \\ \and
		Institute for Structural Mechanics, Ruhr University Bochum, Universit\"atsstra\ss{}e 150, 44801 Bochum, Germany\\ \and
		Department of Structural Mechanics, Gdansk University of Technology, ul. Narutowicza 11/12, 80-233 Gdansk, Poland\\ \and
		Department of Mechanical Engineering, Indian Institute of Technology Guwahati, Assam 781039, India \\ 
		\and Department of Mechanical Engineering,\\
		Indian Institute of Technology Kanpur,
		Uttar Pradesh 208016, India\\
		Corresponding author: \email{\url{pkumar@mae.iith.ac.in}}}
	\maketitle 
	\begin{abstract}
		This paper presents a topology optimization approach to design 2D contact-aided compliant mechanisms (CCMs) that can trace the desired output paths with more than one kink while experiencing self and/or external contacts. Such CCMs can be used as mechanical compliant switches. Hexagonal elements are used to parameterize the design domain. Negative circular masks are employed to remove material beneath them and generate rigid contact surfaces. Each mask is assigned five design variables. The first three decide the location and radius of the mask, whereas the last two determine the presence of the contact surface and its radius. To ensure continuity in contacting surfaces' normal, we employ a boundary smoothing scheme. The augmented Lagrange multiplier method is employed to incorporate self and mutual contact. An objective is formulated using the Fourier shape descriptors with the permitted resource constraint. The hill-climber optimization technique is utilized to update the design variables. An in-house code is developed for the entire process. A CCM is optimized with a two kinks path to demonstrate the method's efficacy. The desired and obtained paths are compared.
		
		\keywords{Topology optimization, Contact-aided Compliant Mechanisms, Self and External Contact, Fourier Shape Descriptor}
		
	\end{abstract}
	\section{Introduction}
	A compliant mechanism (CM) relies on the motion obtained from deformations experienced in its flexible members to perform the desired tasks under an external actuation. Mainly, members' deformations are smooth; thus, output characteristics of CMs are attributed to smooth variations~\cite{kumar2016synthesis,kumar2019computational}. However, in many applications, smooth variations cannot produce the desired tasks, e.g., a path with multiple kinks (non-differentiable paths)~\cite{mankame2002contact,kumar2019computational}. One of the ways to achieve such tasks is to introduce contact constraints in the deformation history of the members~\cite{mankame2002contact,kumar2019computational}. Contact constraints can be generated due to the interaction between members or between members and external bodies~\cite{kumar2019computational}. The former is called self-contact mode, whereas the latter is termed mutual or external contact mode. When a CM uses or requires contact constraints to achieve the desired tasks, such CMs are called contact-aided compliant mechanisms (CCMs)~\cite{mankame2002contact}. These mechanisms can be optimized using topology optimization (TO), an intelligent design optimization technique to achieve optimized material layout within a given design domain~\cite{kumar2016synthesis,kumar2019computational}. There exist various TO approaches; herein, we adopt a feature-based TO method, material masks overlay strategy (MMOS)~\cite{saxena2008material} with contact constraints~\cite{kumar2016synthesis}, for determining the material layout of the CCMs that can trace paths with multiple kinks. Such CCMs can be used as mechanical compliant switches~\cite{nagendra2021topology}.

	Mankame and Ananthasuresh~\cite{mankame2007synthesis} proposed a TO approach to design a non-smooth path (with one kink) generating CCMs wherein the design domain was parameterized using frame elements and the contact location was predefined. The desired and actual curves are compared using the Fourier shape descriptors (FSD) objective~\cite{zahn1972fourier}. Reddy et al.~\cite{nagendra2012systematic} employed curved beam elements to design the domain. Their method determines the contact surface systematically; however, the contact analysis was performed in commercial software. MMOS-based methods are proposed to achieve optimized CCMs tracing non-smooth paths with mutual contact~\cite{kumar2016synthesis,kumar2015synthesis}, self contact~\cite{kumar2017implementation}, and self and mutual contact~~\cite{kumar2019computational,kumar2021topology}. They further extended the method to optimize shape morphing CCMs~\cite{kumar2021topology}. Recently, Reddy and Saxena~\cite{nagendra2021topology} proposed an approach using frame elements to design CCMs that can trace paths with three kinks. The methods permit different contact surface generation and self and mutual contact; however, they use commercial software to perform contact analysis. Inspired by this method and application, we extend our previously presented method in Refs~\cite{kumar2016synthesis,kumar2019computational,kumar2015synthesis} to optimize CCMs that can trace paths with multiple kinks. The entire contact analysis and optimization are performed using code developed in-house. Readers can refer to Refs.~\cite{tummala2014design,jin2020large,huang2023clamping,frederiksen2024topology,navez2024design} for different methods for designing mechanisms with various applications involving contact constraints.
	
	The remainder of the paper is structured as follows. Sec.~\ref{Sec:Sec2_Methodology} provides the presented methodology for designing the CCMs. Sec.~\ref{Sec:sec_3problemformulation} provides problem definition descriptions with boundary smoothing scheme, contact finite element formulation, objective function, optimization formulation, and contact search algorithm. Results and discussions are mentioned in Sec.~\ref{Sec:Sec4_res_dis}. Lastly, conclusions are drawn in Sec.~\ref{Sec:Sec5_Conclusion}.
	
	\section{Methodology}\label{Sec:Sec2_Methodology}
	Material mask overlay method is used wherein the design domain is parameterized using hexagonal elements.   These elements provide edge-to-edge connections between neighboring elements; thus, checkerboard and point connections get automatically subdued from the optimized designs~\cite{saxena2003honeycomb,kumar2023honeytop90}. Negative masks are employed to remove material and generate rigid contact surfaces within them that facilitate external contact/mutual contact. Only self-contact mode will be achieved if only material removal is sought. Mask $m$ is defined via its center coordinate ($x_m,\,y_m$) and radius $r_m$. In addition, ($s_m,\,f_m$) variables are also assigned for generating contact surfaces and corresponding radius. $s_m=1$ and $s_m=0$ indicate rigid contact surface generation with radius $f_mr_m$; whereas latter denotes no contact surface generation. $f_m$ is a fraction number. V-notches on the bounding surfaces are smoothed using the boundary smoothing scheme presented in~\cite{kumar2015topology}; thus, continuity in the normal is maintained, which is important for the convergence of the contact analysis. We use mean shape functions to perform the finite element analysis. The augmented Lagrange multiplier method is employed with the segment-to-segment contact approach. The mechanical state equations are solved using the Newton-Raphson method.
	The primary aim is to generate CCMs that trace the desired paths with multiple kinks. Such CCMs can be used as a mechanical switch application~\cite{nagendra2021topology}. An objective using FSD is formulated to evaluate the differences between the evolving and desired paths. This objective is minimized such that the actual path becomes close to the desired one using a stochastic-based hill-climber search algorithm~\cite{kumar2015synthesis,kumar2016synthesis}. This search selection is stochastic as the design variable $s_m$ for each mask can only take either 0 or 1.
	\section{Problem formulation}\label{Sec:sec_3problemformulation}
	This section briefly notes the boundary smoothing scheme, contact finite element formulation, objective function, optimization formulation, and the hill-climber search algorithm. 
	\subsection{Boundary smoothing scheme}
	Typically, contact analysis is formulated using the interacting boundary/surface's normals; thus, it is desired that normals should be smooth enough for the Newton-Raphson convergence of the contact analysis. As noted, the boundaries of the optimized designs using hexagonal elements are characterized via V-notches\footnote{rectangular elements provide right-angled notches}; thus, jumps in the normals of the boundary normal can be noted~\cite{kumar2016synthesis,kumar2019computational}. These V-notches are subdued using the boundary smoothing scheme presented in~\cite{kumar2015topology} to achieve the desired smoothness. 
	
	To achieve the CCMs with slender members, hexagonal elements and smoothing are removed in two steps. First, the masks remove elements overlaid, and then smoothing is performed. Second, those elements not affected by the smoothing scheme are removed, and boundary smoothing is performed again. This two-step removal also helps control the required volume fraction of the mechanisms while dealing with fewer design variables. 
	
	\subsection{Contact finite element analysis formulation}
	Herein, self and mutual contact modes are permitted. We assume frictionless and adhesionless contact. As we present in our previous study, friction does not play considerable roles in applications with path generation~\cite{kumar2021topology}. Contact is modeled using the augmented Lagrange multiplier method with the Uzawa-type algorithm. Segment-to-segment contact analysis is considered. The classical penalty method is used to determine contact traction in the inner loop of the contact formulation, whereas, in the outer loop, the Lagrange multiplier is updated. The contact traction, $\bm{t}_\text{c}$, is determined in terms of normal gap $g_n=(\bm{x} -\bm{x}_\text{p})\cdot\bm{n}_\text{p}$ as
	\begin{equation} \label{Eq:contacttraction}
		\bm{t}_\text{c} =
		\begin{cases}
			-\epsilon_\text{n} g_\text{n}\bm{n}_\text{p} ~\qquad\text{for}\,\,g_n<0 \\
			\mathbf{0} ~\qquad  ~\qquad \quad \text{for}\,\,g_\text{n}\ge0
		\end{cases}
	\end{equation} 
	where  $\bm{x}_\text{p}$ is the projection point  of $\bm{x}$ on the neighboring contact surface. $\mathbf{n}_p$ is the unit normal at the projection point $\bm{x}_\text{p}$. With this contact formulation, the discretized weak form of the mechanical equilibrium equation in a global statement is 
	\begin{equation} \label{Eq:mechanicalEquilibrium}
		\mathbf{f}(\mathbf{u}) = \mathbf{f}_{\mathrm{int}}+ \mathbf{f}_{\mathrm{c}}-\mathbf{f}_{\mathrm{ext}}  = \mathbf{0},
	\end{equation} 
	where $\mathbf{f}_{\mathrm{int}},\,\mathbf{f}_{\mathrm{c}},\,\text{and}\,\mathbf{f}_{\mathrm{ext}}$ are the internal, contact and external forces, respectively. The elemental internal and contact forces are determined as
	\begin{equation} \label{Eq:elementalinternalforce}
		\begin{split}
			\mathbf{f}^e_\mathrm{int} = \int_{\Omega_{k}^h}\mathbf{B}_\mathrm{UL}^\mathrm{T}\bm{\sigma}\, \mathrm{d}v,\quad 
			\mathbf{f}^e_\mathrm{c} = \int_{\partial \Omega_{k}^h}\mathbf{N}^\mathrm{T}\bm{t}_\text{c}\, \mathrm{d}a,
		\end{split}
	\end{equation}
	where $\bm{\sigma}$ is the Cauchy stress tensor and $\mathbf{B}_\mathrm{UL}$ is the discrete strain-displacement matrix ~\cite{bathe2006finite} of an element in the current configuration. $\mathbf{N} = [N_1\bm{I},\, N_2\bm{I}]$ with $N_1 = \frac{1}{2}(1-\xi), \, N_2 = \frac{1}{2}(1+\xi)$, $\xi\in[-1,\,1]$. Eq.~\eqref{Eq:mechanicalEquilibrium} is solved using the Newton-Raphson method, where a neo-Hookean material model is taken~\cite{kumar2019computational}.
	\subsection{Objective function and optimization formulation}
	We formulate the Fourier shape descriptors objective function that determines the differences between the actual path and the desired one using Fourier shape descriptors. First, a path is closed clockwise without self-intersecting, then its Fourier coefficients are determined.
	
	Let $A_i^k$ and $B_i^k$ be the Fourier coefficients, $\theta^k$ and $L^k$ be the initial orientation and total length of the two curves, $k= a\,,\, d\,$ represent the actual and desired shapes, respectively. $n$ is the total number of Fourier coefficients. FSDs objective is determined as
	\begin{equation} \label{Eq:FSDobj}
		f_0(\bm{\rho}) = w_\text{a} A_\mathrm{{err}}+ w_\text{b} B_\mathrm{{err}}+ w_\text{L} L_\mathrm{{err}}+ w_\mathrm{\theta} \theta_\mathrm{{err}},
	\end{equation}
	where $w_\text{a},\,w_\text{b},\,w_\text{L},\,\text{and}\,w_\theta$ are user-defined weight parameters for the errors 
	\begin{equation} \label{Eq:ErrorDef}
		\begin{aligned}
			A_\mathrm{{err}}&=\sum_{i=1}^{n}(A_i^\text{d}-A_i^\text{a})^2,\qquad
			B_\mathrm{{err}}=\sum_{i=1}^{n}(B_i^\text{d}-B_i^\text{a})^2,\,\\
			L_\mathrm{{err}}&=(L^\text{d}-L^\text{a})^2,\qquad \quad
			\theta_\mathrm{{err}}=(\theta^\text{d}-\theta^\text{a})^2,
		\end{aligned}
	\end{equation}
	The following optimization problem is solved:
	\begin{equation}\label{Eq:optimization}
		\begin{aligned}
			& \underset{\bm{\rho}}{\text{min}}
			& &f(\bm{\rho}) + \lambda_v (V-V^*),\\
			& \text{such that}, & & \mathbf{f}(\mathbf{u}) = \mathbf{0};\,\, \rho_L\le a_i\le \rho_U|_{a_i = x_i,\,y_i,\,r_i}\\ 
			&\,& &s_i\,(= 0\,\, \text{or}\,\, 1)\,;\,\, f_i\,[\in (0,1)]
		\end{aligned}
	\end{equation}
	where $V^*$ and $V^c$ are the desired  and current volumes of CCMs, and $\lambda_v$ is the volume penalization parameter. $\lambda_v = 0$ is set,  when $V^*<V^c$, otherwise $\lambda_v = 20$ is used. $a_L$ and $a_U$ denote the lower and upper limits for $\rho_i\in\bm{\rho}$.
	\subsection{Search algorithm: Hill Climber} 
	We use N$_m$ masks for the optimization process. Mask $m$ is defined by five variables $x_m,\,y_m,\,r_m,\,s_m,\,\text{and}\,f_m$.  A probability parameter $pr\,(=0.08)$ is set for each variable. During each optimization iteration, a random number $\chi$ is generated. If $\chi<pr$, then $\rho_\mathrm{new} = \rho_\mathrm{old} \pm (\kappa\times l)$ is determined, where $0<\kappa<1$ is a generated random number and $l = 0.1\times\max(L_x,\,L_y)$. $s_m$ is updated as follows: if $\chi<pr$ and $\kappa<0.50$, $s_m= 1$, else $s_m=0$. Similarly, $f_m\in[0,\,1]$ is also updated. In addition, the magnitude of the input force is modified as $F_\text{new} = F_\text{old}\pm (\kappa\times l)$ \cite{mankame2007synthesis}. If the input location, output location, and some of the fixed boundary conditions exist in the evolving design, then the FSD objective function is evaluated; otherwise, the design is penalized. Moreover, if $f_\mathrm{new}<f_\mathrm{old}$, the updated design vector is used for the next evaluation. These steps are repeated until the maximum number of iterations is reached, or the process terminates if the change in objective value over ten consecutive iterations is less than $\Delta f = 0.01$.
	
	\section{Results and discussions}\label{Sec:Sec4_res_dis}
\begin{figure}[h!] 
		\begin{subfigure}{0.5\textwidth}
			\centering
			\includegraphics[width=\textwidth]{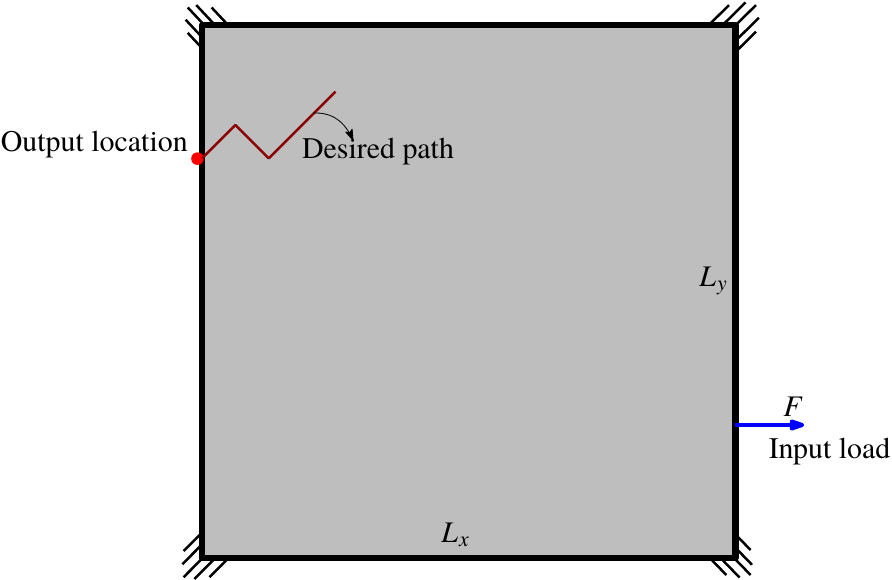}
			\centering
		\end{subfigure}
		\begin{subfigure}{0.45\textwidth}
			\centering
			\includegraphics[width=\textwidth]{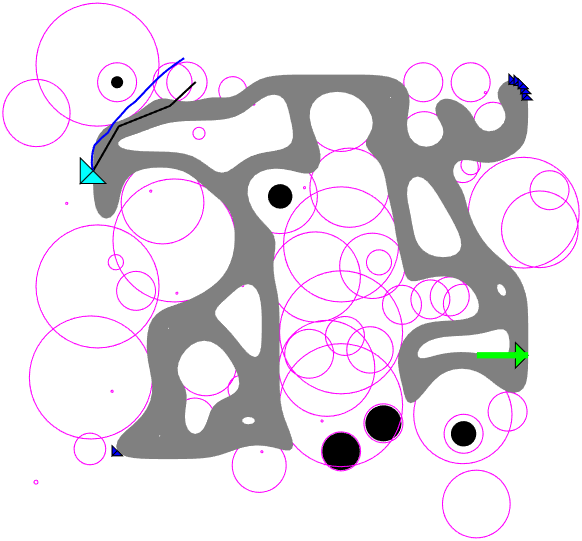}
			\centering
		\end{subfigure}
		\caption{Design domain and the optimized result are shown in (a) and (b). Fixed corners, input location for actuation, and out location with two-kink path are indicated in (a). The optimized design depicts the negative masks' final positions, shapes, and sizes in (b). The optimally generated five rigid surfaces are also depicted using solid black circles.}
		\label{fig:Designdomain}
	\end{figure}

	The presented method is demonstrated by generating a CCM tracing path with two kinks herein. The design domain is depicted in Fig.~\ref{fig:Designdomain}. $L_x$ and $L_y$ indicate dimension in $x$ and $y$ directions, respectively. All the four corners are fixed. The output location with a representative desired path with two kinks is depicted. The input force location with applied force $F$ is shown.  
	
	\begin{figure}
		\begin{subfigure}{0.5\textwidth}
			\centering
			\includegraphics[width=\textwidth]{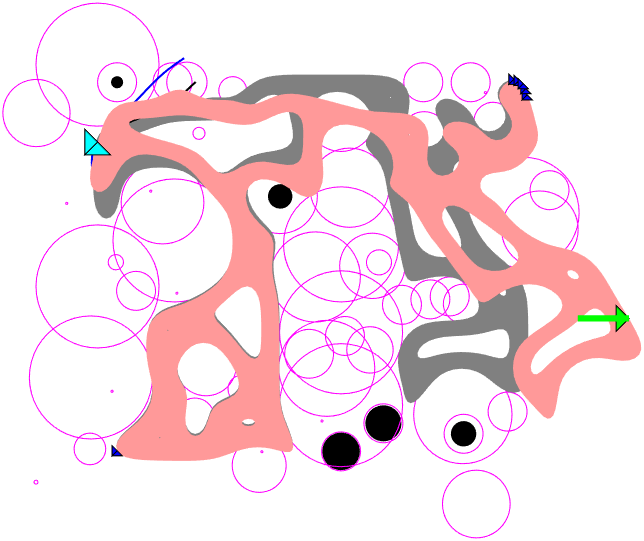}
			\caption{}
		\end{subfigure}
		\begin{subfigure}{0.5\textwidth}
			\centering
			\includegraphics[width=\textwidth]{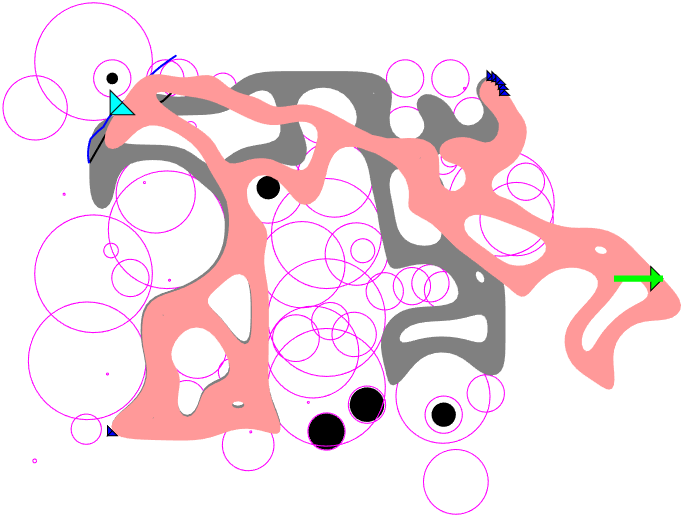}
			\caption{}
		\end{subfigure}
		\begin{subfigure}{0.5\textwidth}
			\centering
			\includegraphics[width=\textwidth]{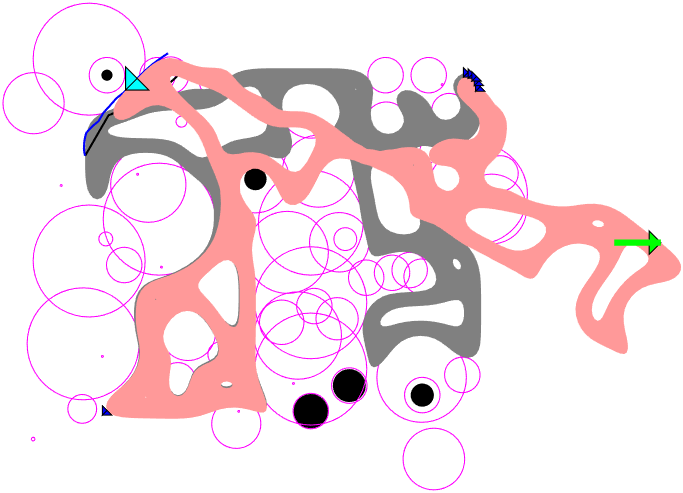}
			\caption{}
		\end{subfigure}
		\begin{subfigure}{0.5\textwidth}
			\centering
			\includegraphics[width=\textwidth]{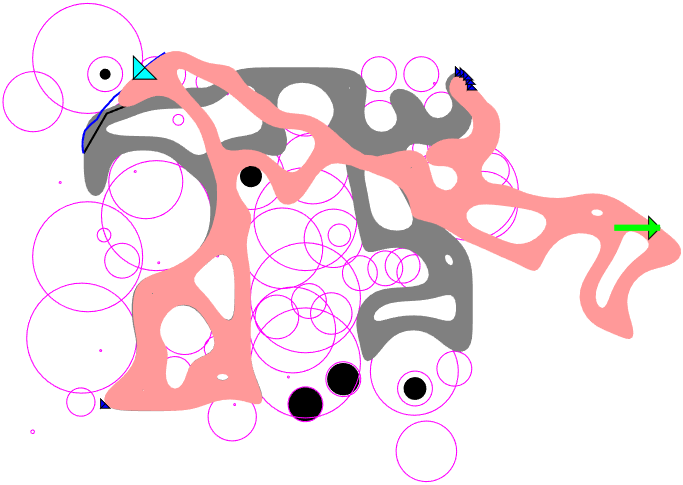}
			\caption{}
		\end{subfigure}
		\begin{subfigure}{0.5\textwidth}
			\centering
			\includegraphics[width=\textwidth]{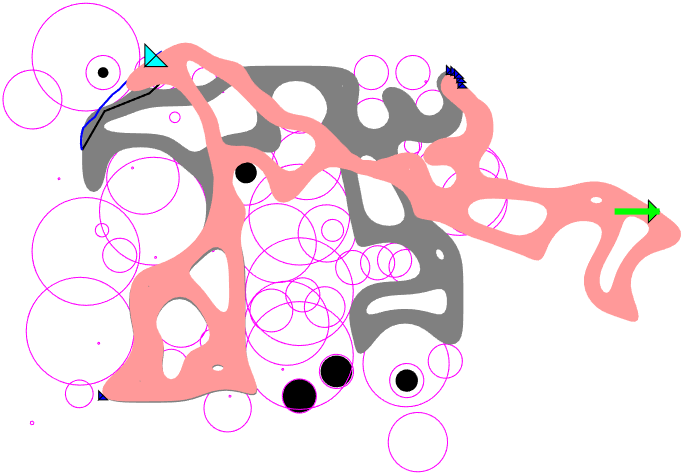}
			\caption{}
		\end{subfigure}
		\begin{subfigure}{0.5\textwidth}
			\centering
			\includegraphics[width=\textwidth]{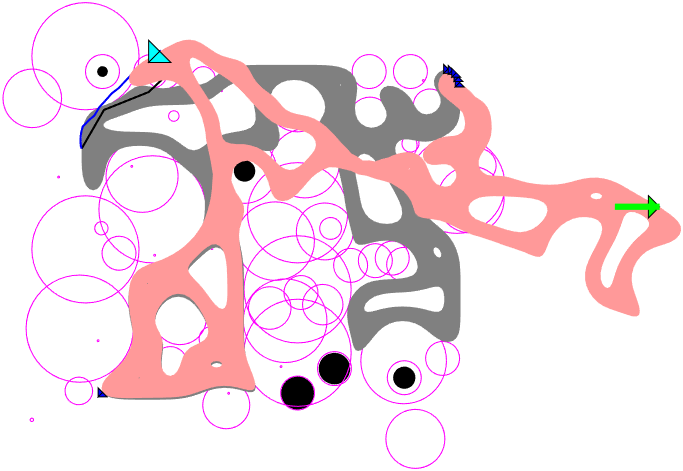}
			\caption{}
		\end{subfigure}
		\caption{Optimized design in gray and deformed profiles in red at different force steps. The desired path and actual trace path are shown in black and blue. The final locations and sizes of the negative circular masks are shown. One can notice that though the optimizer suggests five rigid contact surfaces, the mechanism uses only one of them to perform the task, i.e., only this contact surface is active, while the remaining are inactive; thus, they should be excluded during fabrication.}
		\label{fig:Results}
	\end{figure}
	
	The design domain is parameterized using $30\times 30$ hexagonal elements. We use  \texttt{HexMesher} provided with  \texttt{HoneyTop90} MATLAB code~\cite{kumar2023honeytop90} to generate these elements within the given design domain. The number of masks employed in $x$ and $y$ directions are $N_x =8$ and $N_y =8$. The maximum and minimum radii are set to \SI{6}{\milli \meter} and \SI{0.10}{\milli \meter}. Modulus of elasticity $E =\SI{2.1}{\mega \pascal}$ is used. Poisson's ratio $\nu$ and volume fraction is set to 0.33 and 30\%, respectively. The radius factor for the contact surfaces is fixed to 0.60. Ten steps of boundary smoothing are performed to smoothen the optimized boundary. The upper and lower limits for the applied load are fixed at 1000 N and -1000 N, respectively. Weights $w_a,\,w_b,\,w_L,\,\text{and}\,w_\theta$ are 500, 500, 1, and 0.1, respectively. 
	
	The optimized CCM is depicted in Fig.~\ref{fig:Designdomain}b. The negative masks' final positions, shapes, and sizes are also depicted in Fig.~\ref{fig:Designdomain}b. The results with deformed profiles at different steps are shown in Fig.~\ref{fig:Results} wherein the corresponding positions of applied load and output locations are shown. The desired path and the actual paths are shown in black and blue. The desired path has two kinks. The actual path also has two kinks; however, they are not as sharp as the desired one. One can notice a few small holes are present on the optimized CCM. The optimized CCM has five rigid contact surfaces, however, only one is employed by the mechanism to perform the task. Thus, one should exclude the inactive contact surfaces during fabrication. The holes appear due to the localized removal of one/two FEs. As per~\cite{kumar2021topology}, those holes can be closed for all practical and fabrication purposes as they do not influence the CCM performances. 
	
	Next, we present the comparison between the desired and actual paths. Fourier coefficients in terms of $R_k = \sqrt{A_k^2+ B_k^2}$, where $R_k|_{(k = 1,2,\cdots, n)}$ are curve invariants~\cite{zahn1972fourier}. We evaluate the overall relative change in shape $\zeta_{s}$ as
	\begin{equation}
		\zeta_s = \bigg[\frac{1}{n} \sum_{m=1}^{n} \frac{|R_k^\text{d} - R_k^\text{a}|}{R_k^\text{d}}\bigg],
	\end{equation} 
	where $R_k^\text{d}$ and $R_k^\text{a}$ are invariants corresponding to the desired and actual curves, respectively. Here, we find $\zeta_s=2.398\%$ between these curves (Fig.~\ref{fig:Results}). Likewise, we also determine the relative change in lengths using $\zeta_\text{l} = \frac{|L^\text{d }- L^\text{a}|}{L^d} = 7.26\%$~(Fig.~\ref{fig:Results}). We note that the obtained $\zeta_s$ and $\zeta_l$ are relatively large. This gives us an avenue for future work to develop where we present a more robust method such the CCMs can trace the desired path more closely.
	
	\section{Concluding remarks}\label{Sec:Sec5_Conclusion}
	This paper presents a topology optimization method to design contact-aided compliant mechanisms that trace paths with multiple kinks. This method is an extension of our previously presented methods~\cite{kumar2016synthesis,kumar2019computational,kumar2015synthesis,kumar2021topology} for designing CCMs. The method uses hexagonal elements to discretize the design domain. Negative masks are employed to remove material and generate contact surfaces. Both self and mutual contact modes are permitted; however, the obtained CCM uses only mutual contact modes to achieve the target task. An FSD objective is devised with the allowed volume fraction to compare the desired and actual curves. The optimized CCM traces a path with two kinks. However, the difference between the kinks is not as sharp as in the desired paths. This opens up avenues for future work, where we intend to make the proposed method more robust and experimentally verify the CCM's performance by fabricating it.

	\bibliographystyle{ieeetr}
	\bibliography{Citations}
\end{document}